\documentclass{iau-JDSS}
\usepackage[pdftex]{graphicx}
\usepackage{natbib}

\title[Obscured and distant clusters] 
{SpS5 - I. Obscured and distant clusters}

\author[M.M. Hanson et al.]   
{M.M. Hanson$^1$, D. Froebrich$^2$, F. Martins$^3$, \\ A.-N. Chen\'e$^4$, C. Rosslowe$^5$, A. Herrero$^6$, H.-J. Kim$^7$}

\affiliation{$^1$University of Cincinnati, USA email: hansonmm@uc.edu\\[\affilskip]  $^2$University of Kent, U.K. $^3$LUPM, CNRS \& Universit\'e Montpellier II, France\\[\affilskip] $^4$Universidad de Concepcion/Valparaiso, Chile $^5$University of Sheffield, U.K.,  \\[\affilskip] $^6$Instituto de Astrofisica de Canarias and Universidad de La Laguna, Spain, \\[\affilskip] $^7$Seoul National University, Korea
}

\pubyear{2012}
\volume{Volume 16}  
\pagerange{1--10}
\date{?? and in revised form ??}
\jname{Highlights of Astronomy, Volume 16}
\editors{T. Montmerle, ed.}
\begin{document}

\maketitle

\begin{abstract}
This first  part of Special Session 5 explored the current status of infrared-based observations of obscured and distant stellar clusters in the Milky Way galaxy.  Recent infrared surveys, either serendipitously or using targeted searches, have uncovered a rich population of young and massive clusters.  However, cluster characterization is more challenging as it must be obtained often entirely in the infrared due to high line-of-sight extinction.  Despite this, much is to be gained through the identification and careful analysis of these clusters, as they allow for the early evolution of massive stars to be better constrained.   Further, they act as beacons delineating the Milky Way's structure and as nearby, resolved analogues to the distant unresolved massive clusters studied in distant galaxies.    

\end{abstract}

\firstsection 

\section{Introduction}
Two decades ago, the study of large star forming regions was limited to distant objects, with R136 in the LMC as the closest example.  However, now, several massive young clusters\footnote{Defined as being a few Myr and about 10$^4$\,M$_{\odot}$ in mass - objects with smaller masses would not be detected if distant or obscured, objects with larger masses are few in number.} have been identified in our Galaxy.    First, deep near-infrared studies of the Galactic center region serendipitously identified at least three spectacular and very massive clusters with unique stellar and cluster properties. Then near-infrared surveys, such as the Two Micron All Sky Survey (2MASS) began uncovering massive open clusters deep in the Galaxy.  Numerous new clusters have been found starting with several 2MASS-based targeted near-infrared surveys (Dutra \& Bica 2000, 2001, Froebrich et al\ 2007), GLIMPSE (Mercer et al.\ 2005) and DENIS (Reyle \& Robin 2002).   The search continues with still newer surveys, such as the VISTA Variables in the V\'ia L\'actea (VVV, Borissova et al. 2011).   Moreover, it appears young massive clusters were already known to us, as several long-known optical clusters.  Westerlund 1 \& 2, Stephenson 2, Cygnus OB2 have been found to be far more massive than previously thought (Hanson 2003).

The reasons to study large populations of young clusters are numerous. First, such clusters help us to better reveal galactic structure (Fig.\ 1) and they constrain the history of the chemical evolution throughout the Milky Way. They also provide critical benchmarks to test stellar evolution. For example, having many clusters imply sampling different environments, an essential tool for testing theoretical models. Massive clusters are especially important in this context, as they allow us to beat the small-number statistics that plague typical intermediate-mass cluster studies. With a massive stellar cluster they may display several red supergiants, or several very massive core H-burning stars simultaneously. Finally, the most massive clusters in our Galaxy, and where there is access to study the individual stars, are close analogs to distant starbursts for which only integrated properties are known.  A more in depth knowledge of nearby massive stellar clusters will enhance our understanding of extragalactic stellar clusters.

To improve further our knowledge of distant and obscured clusters, this field will benefit from the forthcoming advent of several new facilities: Gaia, JWST, ALMA, SKA. Besides the detection of new objects, they will provide a better characterization of the identified cases. Notably, an improvement of the distance determinations is awaited, not only for calibration purposes, but also to understand the Galactic structure. Discrepancies have been noted between trigonometric, kinematic and spectrophotometric estimates (Mois\'es et al.\ 2011).  Kinematic values may be wrong if the cluster's velocity is not only fixed by the global kinematic field of the Milky Way ({\em e.g.} if the cluster formation results from the collision of two moving clouds).

\section{The search and characterization of obscured and distant young stellar clusters}

Young and massive clusters of stars are rare, distant and often obscured by surrounding gas and dust. However, their energy and momentum input into the ISM via e.g. stellar winds, ionizing radiation or SN explosions is extensive. A first step to understand the details of the evolution and feedback from massive stars is to find and characterize their birth clusters.

\begin{figure}[]
   \centering
   \includegraphics[width=4in]{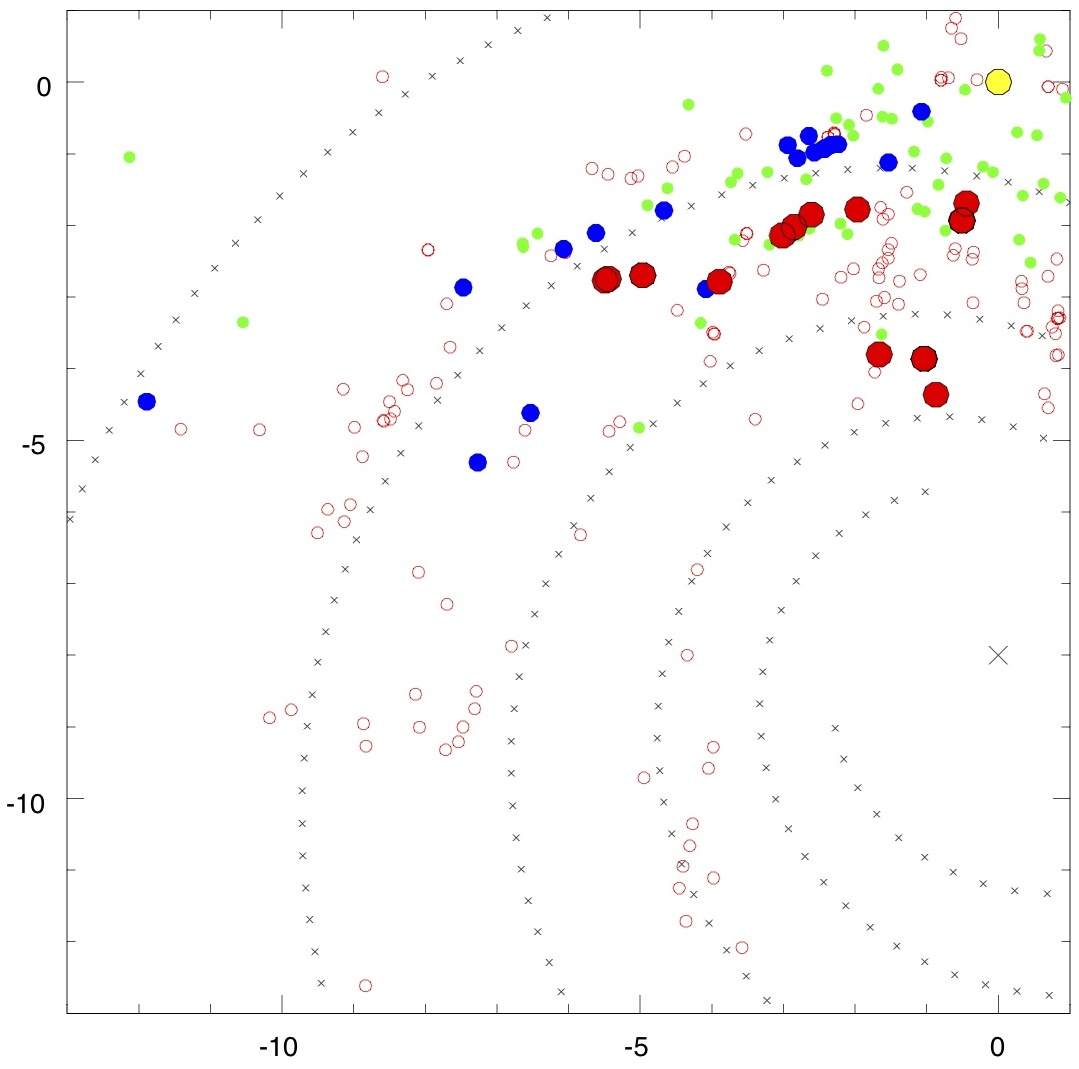} 
   \caption{{\bf Infrared clusters as spiral arm tracers}.  Vall\'ee (2008) model for the Milky Way spiral arms is given in x's, the Sun is at [0,0].   Small open red circles mark HII tracers, small green small filled circles mark Cepheid distances.  The larger, filled blue and even larger filled red dots represent young clusters, optical and infrared, respectively, being studied by the VVV survey (figure from Chen\'e et al.\ 2012). }
   \label{fig:example}
\end{figure}

Infrared-based searches for obscured and distant massive clusters in the Milky Way has progressed greatly since the first 2MASS cluster searches.  Higher spatial resolution, deeper photometry, and numerous new algorithms have been obtained and applied to discover ever more candidate stellar clusters.  With so many searches underway, will the numbers ever reach the 30,000 stellar clusters in the Milky Way as predicted by Portegies Zwart et al.\  (2010)?   Currently, the cluster mass function in the Milky Way is underrepresented in young massive clusters older than a few tens of millions of years and less than 100 million years.  This has driven a boom of stellar cluster searches in the Milky Way using recent, current and future surveys such as 2MASS, Spitzer, WISE, VISTA.  They provide outstanding, homogenous large-scale data sets and are ideally suited for such searches.  

Unbiased, targeted and serendipitous methods have already been mentioned as a means for locating new young clusters.  Also, many``special" targets point towards massive cluster candidates:  $\gamma$-ray or X-ray sources, and Luminous Blue Variables (LBV) or Wolf-Rayet (WR) Stars might be identified and are typically located in or near a massive young stellar cluster (Homeier et al.\ 2003).   Locating any kind of stellar remnant or highly massive star provides a high likelihood of locating a massive young cluster nearby.  

Despite the abundance search methods, targeted, simple visual (by eye) searches, sophisticated algorithms searching for increased stellar densities or photometric color ranges, no method can be sure to catch all clusters of a certain distance or luminosity (or mass) limit and each has its own biases.  Variations, predominantly in extinction, but also the complexity of the background star field, the concentration level of the stellar cluster, and the age and  mass of the cluster will strongly influence the detectability of real clusters (Hanson et al.\ 2010).  This makes the derivation of such values as the cluster luminosity function and the cluster destruction timescales nearly impossible to obtain confidently for the Milky Way.  A number of cluster databases now exist on the web (WEBDA, Dias, Kharchenko, SAI Catalog, FSR listing, etc.), yet they represent just 10\% of what is expected to exist in the Milky Way. Moreover, false positives are expected to be rather high with the deepest searches.  

Many of the same methods used for obtaining age, distance and extinction of optical clusters are applied in the infrared for these clusters with some success: color-magnitude diagrams, spectral classification, etc ({\em e.g.} Borissova et al.\ 2011, Chen\'e et al.\ 2012b).  However, characterization of infrared clusters is more challenging than in the optical due to the weaker dependence of infrared colors with temperature variation and the strong contamination by red-giant field stars.  For clusters in the age range of a few to tens of millions of years, their most massive stars may exist as exceedingly bright, red supergiants.  Their blaring luminosity impedes study of the rest of the cluster (including identification of the cluster in the first place).   Radial velocities combined with galactic rotation curves can provide some independent values for distance (though see Mois\'es et al.\ 2011).  Maser parallaxes are most promising.  In massive young clusters, there are sure to be red supergiants and these can provide strong radio masers, useful for parallax and proper motion studies to derive distance and cluster membership (Zhang et al.\ 2012).  

Cluster mass is always a difficult characteristic to determine as it is strongly dependent on assumptions of the initial mass function, lower mass cutoff and binary fraction.   Virial masses can sometimes provide better insight, but not many infrared clusters have been studied in this way, due to the lack of high resolution spectra of constituent stars. Virial mass has been used as a mass measure with integrated spectra of extragalactic clusters (Mengel et al.\ 2002).  Promising future surveys include exploring the time domain ({\em e.g.} VVV and LSST) for variability and proper motions, as well as narrow band surveys to locate variable (LBVs) or emission-line stars (WRs) expected to be associated with massive stellar clusters.

\section{Massive stars in the Galactic Center}

The center of the Galaxy is a unique environment to study massive stars. On a scale of about 100 parsecs, three young massive clusters are encountered: the Arches, the Quintuplet and the central cluster. In addition to the massive clusters, a number of ``isolated'' massive stars are also found. In total, about 90 Wolf--Rayet stars and more than 250 OB stars have been identified in the last 20 years. The Galactic Center is hidden behind 25 to 30 magnitudes of visual extinction, rendering optical observations unfeasible. Hence, it is perfectly suited for infrared studies. In addition, its distance is well constrained: 8$\pm$0.5 kpc (Gillessen et al.\ 2009). This is important since it allows good luminosity determinations.

{\bf The Arches cluster:} It was first discovered by Nagata et al.\ (1995) and  Cotera et al.\ (1996). It is the youngest and also the most compact of the three clusters. The current census of the massive star population includes 13 Wolf--Rayet stars and more than 100 OB stars.  All Wolf--Rayet stars are late--type WNh stars. There is no WC or red supergiant. The quantitative analysis of their fundamental properties, as well as of the brightest O supergiants, confirms that the population is young: 2--3 Myr (Figer et al.\ 2002, Martins et al.\ 2008). The analysis of the H, He, C and N abundances of the WNh stars indicates that they are still core--hydrogen burning objects. They are very luminous. Some of them might have masses in excess of 120 M$_{\odot}$ (Martins et al.\ 2008, Crowther et al. 2010). The nitrogen content of the WNh stars is a metallicity indicator (Najarro et al.\ 2004). A value of Z=1.2--1.4 Z$_{\odot}$ is preferred. The present-day mass function is still debated: early studies by Figer et al.\ 1999 and Stolte et al.\ (2005) indicated a top--heavy mass function ($\Gamma = -0.7..-0.9$), but the recent analysis by Espinoza et al.\ (2009) points to a value almost consistent with the Salpeter slope ($\Gamma = -1.1\pm 0.2$). What is clear from all studies is that the cluster is mass segregated: the most massive stars are located in the central region. This segregation could be due to dynamical interactions (Kim et al.\ 2006, Harfst et al.\ 2010) or could be primordial (Dib et al.\ 2007).   

{\bf The Quintuplet cluster:} The stellar content comprises 21 Wolf--Rayet stars (13 WN + 8 WC) and about 85 OB stars (Liermann et al.\ 2012).  Several WN stars are the same type of objects as the Arches WNh stars: very luminous with a high hydrogen content (Leirmann et al.\ 2010). The analysis of the OB population and of WR/O ratios indicate an age of 4$\pm$1 Myr (Liermann et al.\ 2012), slightly older than the Arches. Hu{\ss}man et al.\ (2012) derived the mass function and found a slope slightly steeper than the Salpeter mass function ($\Gamma = -0.68\pm 0.13$). The Quintuplet clusters hosts a number of peculiar objects:  Tuthill et al.\ (2006) showed that at least two of the five WC stars from which the cluster was named are binaries showing ``pinwheel nebulae''; the Pistol star, a Luminous Blue Variable, is also located very close to the cluster. Once thought to be the most massive star in the Galaxy, it is now considered to be a binary (Martayan et al.\ 2011) with the most massive component of which being a $\sim$100 M$_{\odot}$ star.  Detailed analysis of its infrared spectrum shows that it has a solar Fe content, and $\alpha$ elements abundances larger by a factor of two compared to the Sun (Najarro et al.\ 2009).

{\bf The Central cluster:} It hosts the supermassive black hole at the position of the radio source SgrA*. The current census of the massive star population includes 31 Wolf--Rayet stars (18 WN + 13 WC) and more than 140 OB stars. Three red supergiants are also located in the cluster. Based on the positions of OB stars in the HR diagram, Paumard et al.\ (2006) estimated an age of 6$\pm$2 Myr. The detailed analysis of the physical properties of most of the Wolf--Rayet population allowed Martins et al.\ (2007) to define the following evolutionary sequence for stars with masses $\sim$ 50 M$_{\odot}$:

\hskip0.8in  (Ofpe/WN9 $\leftrightarrow$ LBV) $\rightarrow$ WN8 $\rightarrow$ WN8/WC9  $\rightarrow$ WC9

The mass function has been analyzed by Bartko et al.\ (2010): it is extremely top--heavy between 1'' and 12'' from the position of SgrA*, and becomes consistent with Salpeter beyond. This is consistent with the simulations of star formation in a gas cloud falling in the Galactic Center (Bonnell \& Rice 2008). Binaries are present in the cluster (Martins et al.\ 2006) and can explain the spectral type of the bow--shock star IRS8, much earlier than any other O star in the cluster (Geballe et al.\ 2006)

{\bf Isolated massive stars:} Two main techniques have been used to identify massive stars outside the three clusters: P$\alpha$ narrow-band imaging (Cotera et al.\ 1999, Homeier et al.\ 2003, Mauerhann et al.\ 2010) and cross--correlation between Chandra and 2MASS catalogs (Muno et al.\ 2006, Mikles et al.\ 2006). Follow-up spectroscopy has revealed the presence of Wolf--Rayet and OB stars of all types. Currently, 26 WR stars (16 WN + 10 WC) are known outside the clusters. They might have been ejected from those clusters by dynamical interactions. But they might as well result from a constant, low efficiency star formation in the Galactic Center, on top of which extreme starbursts events leading to massive clusters happen from time to time.  The Galactic Center is a unique environment for the infrared studies of massive stars properties, formation and evolution.

\section{Young stellar clusters in the VVV Survey}

VISTA Variables in the V\'ia L\'actea (VVV) is one of the six ESO Public Surveys selected to operate with the new 4-meter Visible and Infrared Survey Telescope for Astronomy (VISTA).  It is currently performing unprecedented deep infrared observations of the Galaxy's bulge and an adjacent section of the mid-plane. This unique database is being used to achieve a large (700-800, including some discovered by us in VVV data; Borissova et al. 2011, Bonatto et al. in prep), statistically significant sample of star clusters with homogeneously derived (i.e. all observed with the same instrument and set-up) physical parameters (including angular sizes, radial velocities, reddening, distances, masses, and ages).  This sample will represent a real gold mine for testing and constraining the theories of star cluster formation and evolution. 

The first phase of the VVV survey was completed last year and deep near infrared images in all the filters of the whole area covered by the survey have been obtained.  These data and the analysis methods have been presented in a recent paper (Chen\'e et al. 2012a). For all the clusters and cluster candidates included in the survey statistically decontaminated and analyzed the color-magnitude diagrams (CMDs) have been calculated.  A spectroscopic follow-up has been conducted to confirm their cluster nature, and to derive the spectral types and distances of the brighter cluster members. 

As a first step, efforts were focused on young open clusters in their first few Myrs. During this period, which corresponds to Phase I in the recent classification of Portegies Zwart et al.\ (2010), stars are still forming and the cluster contains a significant amount of gas. The cluster evolution during this phase is governed by a complex mixture of gas dynamics, stellar dynamics, stellar evolution, and radiative transfer, and is not completely understood (Elmegreen 2007, Price \& Bate 2009). Thus many basic (and critical) cluster properties, such as the duration and efficiency of the star-formation process, the cluster survival probability and the stellar mass function at the beginning of the next phase are uncertain. A subsample of fairly massive young clusters have been recently studied and presented by Chen\'e et al. (2012a, 2012b) and Baume et al. (in prep). On a hybrid map of of the Galaxy, plotting this subsample in combination with data from previous papers, it shows that clusters detected in the optical regime form a group offset and parallel to the group of clusters studied in the IR based on VVV data (Fig.\ 1).  In fact, both traces may reveal the internal structure of only one arm (the Carina arm), as can be seen in other galaxies (e.g. M51). Clusters discovered in the VVV data would be embedded in dark clouds behind the young (optical) populations at the front end of the arm.

\begin{figure}[t]
   \centering
   \includegraphics[width=1.7in]{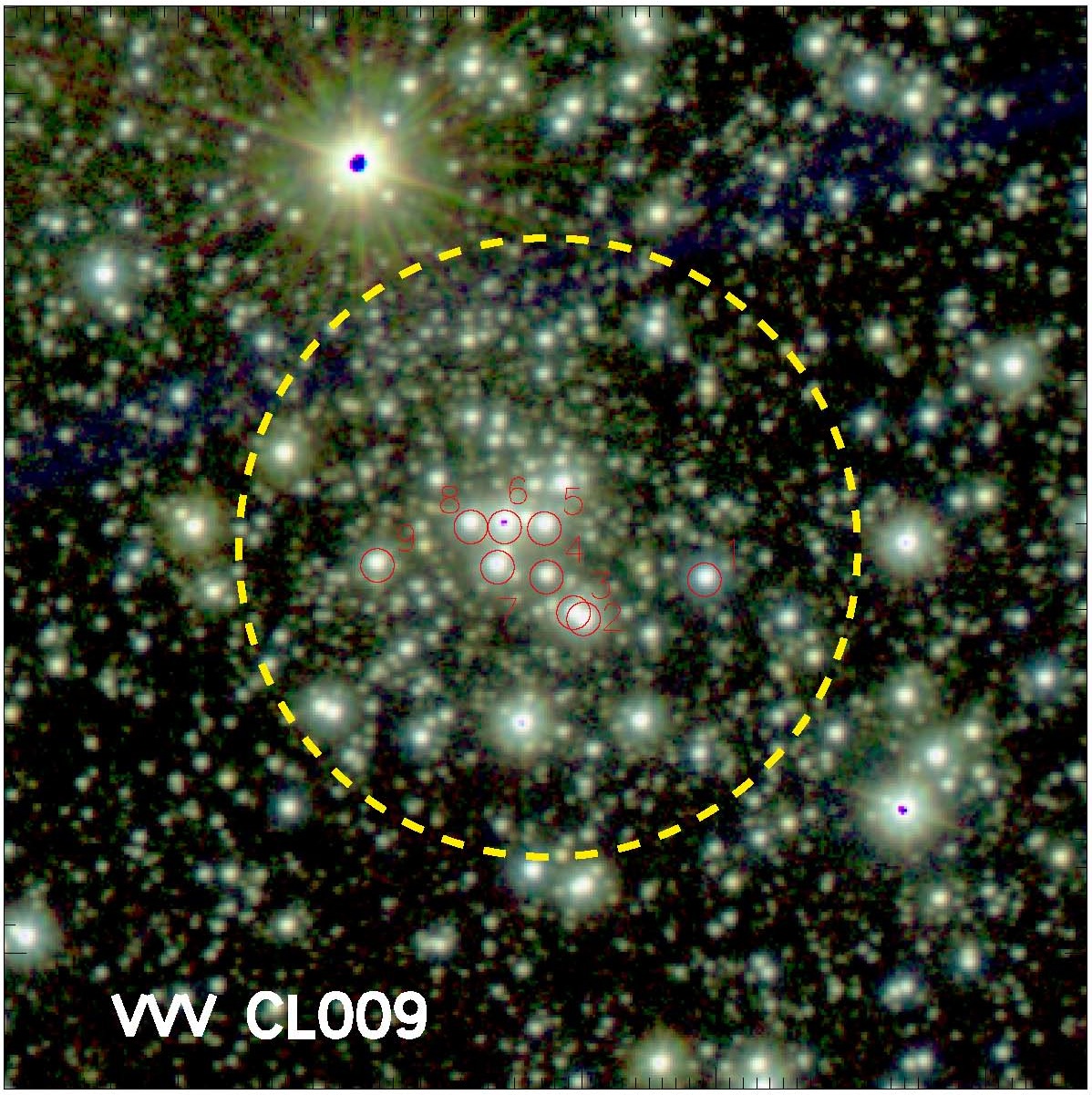} 
      \includegraphics[width=1.7in]{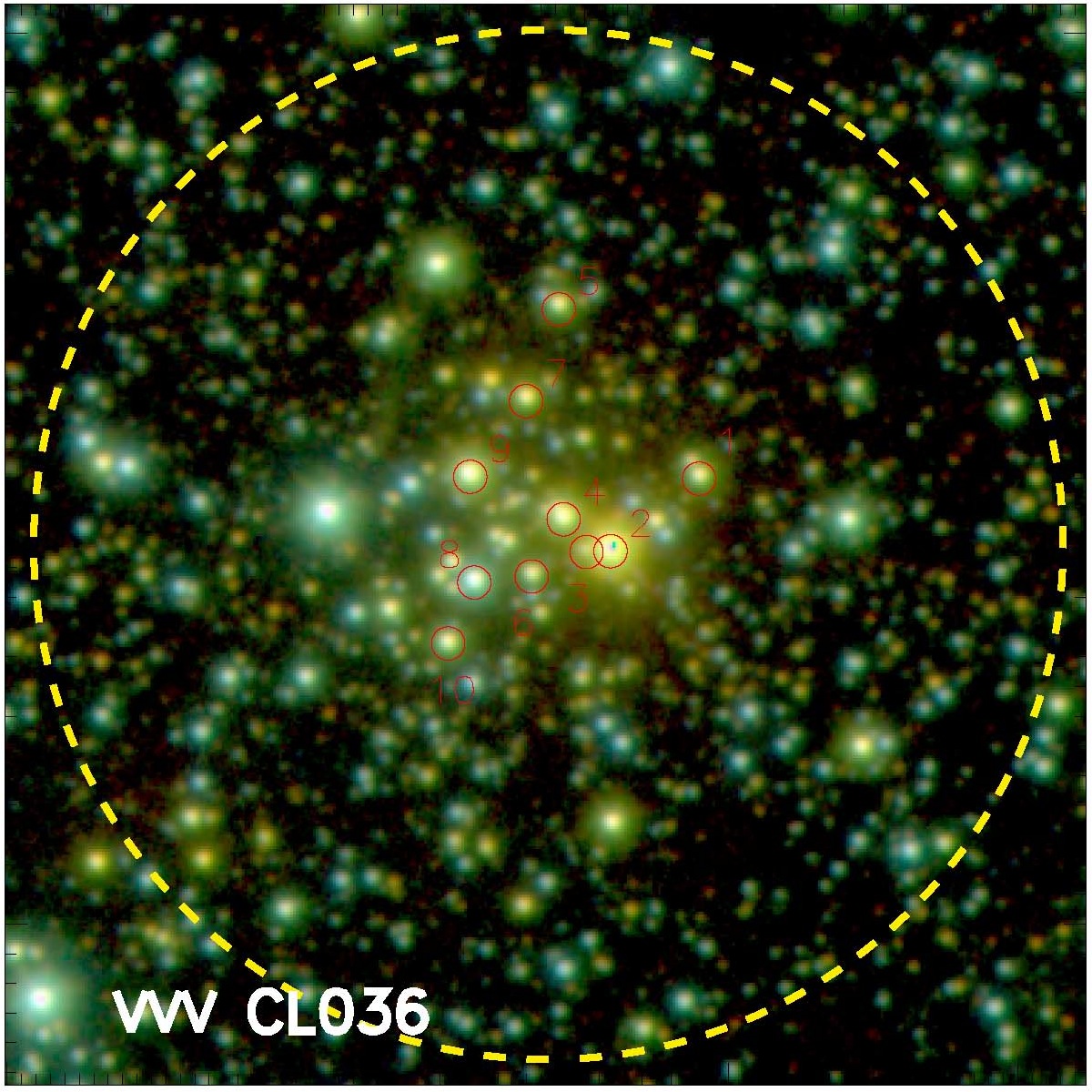} 
         \includegraphics[width=1.7in]{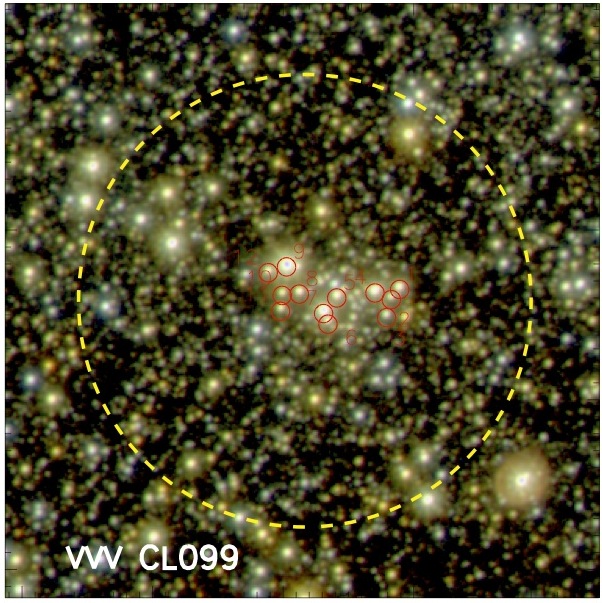} 
   \caption{New infrared stellar clusters, identified by the VVV Survey (Chen\'e et al.\  2012b).}
   \label{fig:example}
\end{figure}

\section{Distribution of WR stars in the Milky Way from near-IR surveys}

The past two decades have seen a three-fold increase in the number of known Galactic WR stars. As the majority of these newly discovered objects are visually obscured, it is essential to develop our capability to analyze these rare and valuable stars at IR wavelengths. Furthermore, a comprehensive knowledge of the Galactic WR distribution will facilitate important tests of massive star evolution theory.   A near-IR classification scheme has been developed for Nitrogen (WN) and Carbon (WC) sequence \textrm{WR} stars from \textrm{IRTF SpeX} $1$-$5\mu$ m spectra of $29$ robustly (optically) classified \textrm{WR} stars. The resulting scheme is based primarily on emission line equivalent width ratios throughout this spectral range. For \textrm{WN} stars, an exact spectral type is achievable with a medium/low resolution $1$-$5 \mu$ m spectrum. An accuracy of $\pm1$ is achievable with a $J$- or $K$-band spectrum alone, and $\pm2$ is achievable with only a $H$- or $L$-band spectrum. For \textrm{WC} stars, it is possible to obtain an exact spectral type based on a medium/low resolution $J-$band spectrum only. Accuracies of $\pm1$ and $2$ are achievable using $K$-band and $H$- or $L$-band spectra, respectively.

Using this new near-IR classification capability, revised spectral types of a large number of recently reported WR discoveries have been obtained. With accurate spectral types in place, an absolute magnitude-subtype calibration has been performed comprising of $83$ and $22$ Galactic \textrm{WN} and non-dusty \textrm{WC} (figure $3$) stars with distance measurements taken from the literature. The extent of reddening to each object in this calibration sample is calculated by a color excess method invoking the intrinsic \textrm{WR} colours of Crowther et al. (2006). As can be seen in figure $3$, a factor of $\sim 5$ increase in the \textrm{WR} sample size over Crowther et al. (2006) provides, for the first time, an absolute magnitude value for each individual \textrm{WN} and \textrm{WC} subtype.  

\begin{figure}[htbp]
   \centering
   \includegraphics[width=2.63in]{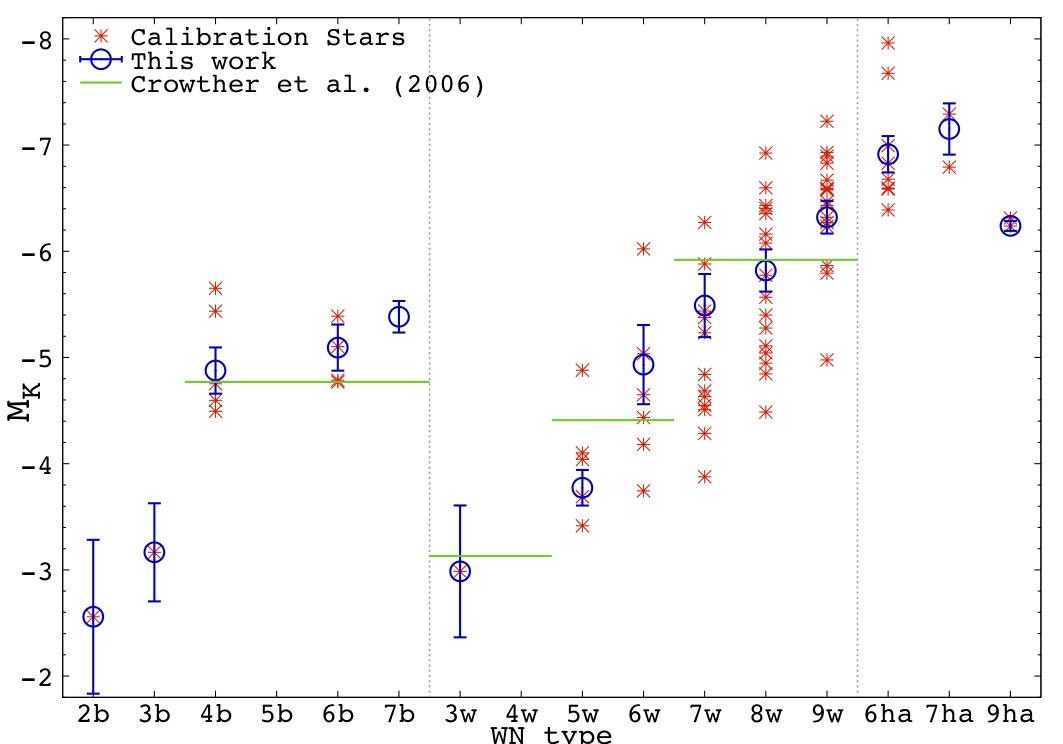} 
   \includegraphics[width=2.63in]{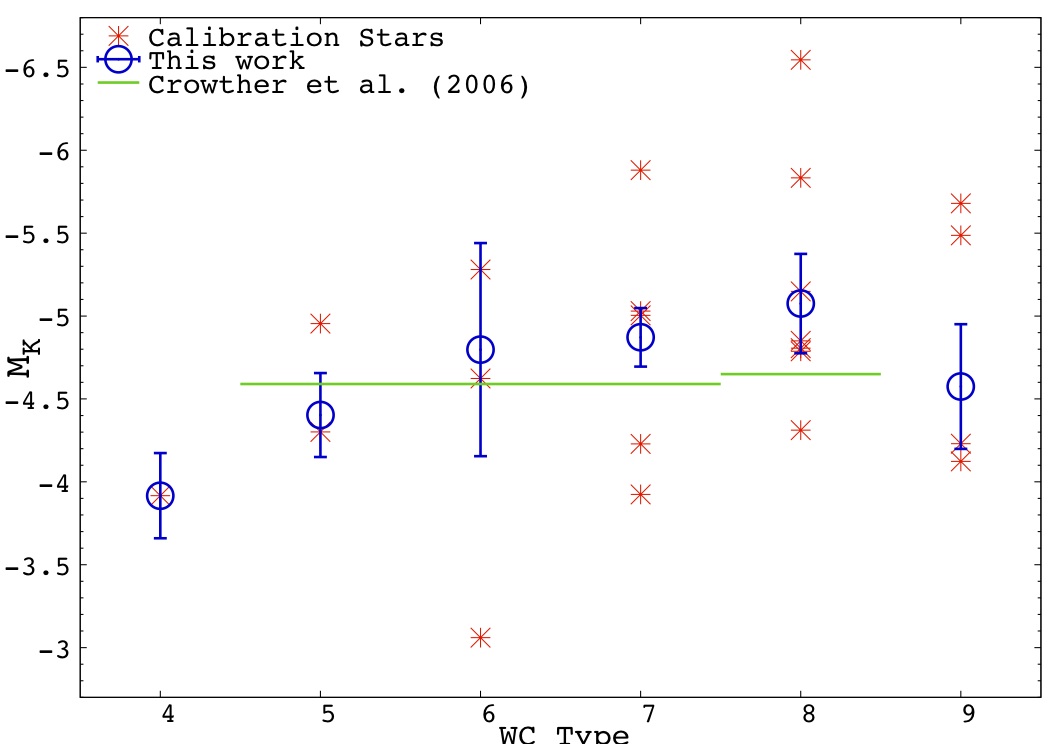} 
   \caption{Absolute magnitude-subtype calibration of 83 WN and 22 WC Wolf-Rayet stars with distance measurements from the literature}
   \label{fig:example}
\end{figure}

The availability of reliable $M_K$ values for each \textrm{WR} subtype has allowed us to map the spatial distribution of the known Galactic \textrm{WR} population.  So far this has been accomplished for $332$ Galactic Wolf-Rayet stars which appear to be without a massive companion.  With knowledge of the Galactic metallicity gradient, this population of $332$ \textrm{WR} stars has been divided into sub-solar ($\log(\frac{O}{H})+12 \simeq 8.5$), solar ($\log(\frac{O}{H})+12 \simeq 8.7$) and super-solar ($\log(\frac{O}{H})+12 \simeq 8.9$) metallicity regions. Comparing the \textrm{WC}:\textrm{WN} number ratios across this metallicity range to the predictions of non-rotating evolutionary models by Eldridge \& Vink (2006) and rotating evolutionary models by Maeder \& Meynet (2005), shows that at solar metallicity and above, the predicted \textrm{WC}:\textrm{WN} number ratio is too low in the rotating case and too high in the non-rotating case. At this time, the number of \textrm{WR} stars in the sub-solar Galactic population is too low to provide a meaningful comparison to evolutionary predictions. A possible interpretation of the discordance of our number ratios with evolutionary predictions will be attempted after the inclusion of further Galactic \textrm{WR} stars, specifically those in binary systems.     

\section{Massive clusters in the Milky Way}

The number of massive young stellar clusters (M$_{cluster} \ge 10^4$ M$_{\odot}$) known in the 
Milky Way has dramatically increased in the last years thanks to recent advances in NIR
observations and the exploitation of large NIR photometric surveys. Rough estimates,
however, indicate that the actual number of massive young stellar clusters may exceed 
by at least a factor of five the number of known ones (Hanson \& Popescu 2008). 
And the number of intermediate mass clusters is much larger.

The MASGOMAS ({\bf MA}ssive {\bf S}tars in {\bf G}alactic {\bf O}bscured {\bf MA}ssive cluster{\bf S}),  
a project aimed at studying the stars in these clusters (see Mar\'in-Franch 
et al.\  2009) was started.  this effort was begin with a search of massive young obscured clusters in a limited region of 
the Milky Way using photometric cuts optimized for the search of OB-stars 
(Ram\'irez-Alegr\'ia, Mar\'in-Franch \& Herrero, 2012). Moreover, the
search was limited to stars with Ks$<$12.5 to allow a spectroscopic follow-up with LIRIS@GTC.

In this way, two new clusters have been found in the Milky Way. The first, MASGOMAS-1
(Ram\'irez-Alegr\'ia et al.\ 2012), was identified
as candidate by the presence of a concentration of candidate OB-stars in the
photometric diagrams and was confirmed by the presence of
17 stars classified as O9-B1V, 3 M supergiants and one A supergiant among the 28 
stars observed spectroscopically in the follow-up programme 
(of course, this is a minimum for the cluster population). 
Based on this spectroscopically confirmed population, a
distance to MASGOMAS-1 was found of 3.53+/-1.48 kpc, with a mass of $1.94 \times 10^4 $M$_{\odot}$ (using 
the IMF from Kroupa 2001) and an age between 6.5 and 10 Myr, based on the
simultaneous presence of red supergiants and O9V stars.

The second cluster, MASGOMAS-4, has two cores surrounded by strong nebulosity at 5.8~$\mu$m.  One of the cores is the HII region
Sh2-76 E, known as an active massive star forming region (Wu et al.\ 2007). 
It is unclear whether the two cores are physically related.  Spectroscopic studies of the stellar population of both cores has found that they
are at the same distance from the Sun and have the same average extinction. This indicates that both cores are physically related, and that the whole cluster, at
a distance of 1.9 kpc, has a mass of $2.19 \times 10^3 $M$_{\odot}$, and an age $<$ 3 Myr, based on the presence of the HII region. 
However, there is a significant difference between both cores: core B displays in its color-color diagram a 
population of young, possible pre-main sequence objects, while this population is
absent in core A. The reason for this difference is not yet known.

\section{Massive stars with infrared excesses in G54.1+0.3}
G54.1+0.3 is a young ($\sim$3,000 yr) Crab-like supernova remnant (SNR) that hosts numerous point-like, strong infrared-excess sources in its outer IR loop.  Color-color diagrams reveal a cluster of possibly hot, massive stars, with $A_V = 7 - 9$.  Spectral energy distributions (SED) of these sources show large mid-infrared excesses, with a strong dip in emission between 6 $-$ 10 $\mu$m.  Infrared classification spectroscopy indicates the brightest are late-O and early-B, with a distance of about 6 kpc.  Did the supernova trigger this star formation?  Models of the SED indicate the dust is located at a distance from the stars, probably originating from the SN, indicating a causal connection.

\section{The Scutum Complex}

The region of the Galactic Plane between $l$=24 and $l$=30 contains a number of clusters of
red supergiants, which are believed to be very massive. Neg have developed techniques to
identify red supergiants from infrared photometric catalogues and used them to search for
such objects over a wide field.  Hundreds of red supergiants have  been found with radial velocities close
to the terminal value corresponding to the Scutum tangent.  Their distribution has been analyzed to characterize this vast region of star formation, known as the Scutum Complex.   Red supergiants in such numbers can be used to trace Galactic structure better
than any other usual tracer. Finally, numerous methods are being explored to detect and observe red supergiants through high extinction.

\section{Finding Red Supergiants in the Galaxy and Measuring their Distances}

Red supergiants evolve from moderate to fairly high-mass stars. They are thus tracing recently formed, high-mass clusters.   They are extremely bright in the infrared, and serve as beacons amongst the dimmer main-sequence objects in a heavily extinguished cluster.  Most uniquely, they have the potential to be naturally strong radio masers, created by simple molecules in their outer circumstellar envelopes (Chapman \& Cohen 1986).   These masers make excellent kinematic probes, as their line-of-sight velocities follow directly from the centroid of the line profiles, revealing kinematic distances when used with a rotation curve.  Moreover, very long baseline interferometric observations can provide direct determination of distances and proper motions (Messineo et al.\ 2002).  More sophisticated methods for locating these very useful objects in the Milky Way is well underway. They use $JHK$ colors and look for the telltale sign of an extended circumstellar shell where a radio maser might be detected (Messineo et al.\ 2012).  

\section{Poster contributions to this session}

A.\ Bonanos presented new results from an ongoing survey to discover and characterize massive binaries in young massive clusters through out the Milky Way, with targeted efforts in Westerlund 1, the Danks and Arches clusters.   Massive binaries are important to constrain theoretical models of star formation and evolution in single and binary stars.

A.\ Damineli presented a near-infrared study of the stellar content of 35 HII regions in the galactic plan, 24 of which are giant HII regions.  Their goal is to confirm, using color-magnitude diagrams, if the kinematic distances seem consistent with the stellar luminosities.  About half are in agreement, but the remaining half appear to be closer than their kinematic distance indicates (Mois\'es et al.\ 2011).  Several are further confirmed to be closer than their kinematic distances based on additional spectrophotometric parallaxes.   

T.\ Geballe presented new $J-$ and $H-$band spectra of several of the bright stars in the young, very massive galactic center Quintuplet cluster.  This allowed the photospheres to be clearly seen for the first time, as longer wavelength spectra had been dominated by emission from their dusty, warm cocoons.  GCS3-1, GCS3-4 and GCS4 all resemble late-type WC stars.  $H-$band spectra of GCS3-2 reveal lines of He I and C II initially in 2009, but then missing in 2011, possibly due to increased dust production.  

C.C.\ Lin presented  a star-counting algorithm to search the 2MASS point source catalog for density enhancements indicative of new stellar cluster candidates.  One such enhancement located the open cluster G144.9+0.4, which they went further to analyze using proper motions to constrain membership.  Several classical T-Tauri stars were found to be members of the cluster.  Along with fits to their spectral energy distribution, these indicate a very young age for the cluster. 

N. Panwar presented a multi-wavelength study of the HII region IC~1805.  They analyzed the spatial extend and structure of the young stellar objects, identified by their spectral energy distributions.  These were found to be mainly intermediate-high mass stars.

\section{Conclusions}
The identification and characterization of obscured, distant massive clusters is still in its early phases, marked by an initial excitement over the awareness of the existence of young massive clusters being in and an important component of the Milky Way.   How might these clusters help us constrain the structure and evolutionary history of the Milky Way?   What critical questions do Milky Way massive clusters uniquely answer to increase our understanding of the formation and evolution of massive stellar clusters in {\em all galaxies}?  Now is the time for the community to assess how the upcoming new facilities, combined with our own innovative ideas might be applied in the next decade to make this a very promising research direction.


\end{document}